\documentclass{article}
\usepackage{IEEEtrantools}
\usepackage{spconf,amsmath,graphicx}
\usepackage{latexsym}
\usepackage{amssymb}
\usepackage{bm}
\usepackage{multirow}
\usepackage{xcolor}
\usepackage{lipsum}
\usepackage{multicol}
\usepackage{enumerate}
\usepackage{algorithm}
\usepackage{algorithmicx}
\usepackage{algpseudocode}
\usepackage{subfig}
\usepackage[utf8]{inputenc}
\usepackage{float}

\ninept
\title{NEAR-OPTIMAL INTERFERENCE EXPLOITATION 1-BIT MASSIVE MIMO PRECODING VIA PARTIAL BRANCH-AND-BOUND}
\name{Ang Li$^*$, Fan Liu$^\dag$, Christos Masouros$^\dag$, Yonghui Li$^*$, and Branka Vucetic$^*$
\thanks{This work was supported in part by the European Union's Horizon 2020 research and innovation programme under the Marie Sk\l{}odowska-Curie Grant Agreement No. 793345, in part by the Engineering and Physical Sciences Research Council (EPSRC) project EP/R007934/1, in part by the Science and Technology Program of Shaanxi Province under Grant No. 2019KW-007, in part by the Australian Research Council (ARC) under Grant DP150104019 and Grant DP190101988, and in part by the ARC Laureate Fellowship under Grant FL160100032.}}
\address{$^*$School of Electrical and Information Engineering, University of Sydney, NSW 2006, Australia.\\
$^\dag$Department of Electronic and Electrical Engineering, University College London, WC1E 7JE, UK.}
\begin{document}
%
\maketitle
\begin{abstract}
In this paper, we focus on 1-bit precoding for large-scale antenna systems in the downlink based on the concept of constructive interference (CI). By formulating the optimization problem that aims to maximize the CI effect subject to the 1-bit constraint on the transmit signals, we mathematically prove that, when relaxing the 1-bit constraint, the majority of the obtained transmit signals already satisfy the 1-bit constraint. Based on this important observation, we propose a 1-bit precoding method via a partial branch-and-bound (P-BB) approach, where the BB procedure is only performed for the entries that do not comply with the 1-bit constraint. The proposed P-BB enables the use of the BB framework in large-scale antenna scenarios, which was not applicable due to its prohibitive complexity. Numerical results demonstrate a near-optimal error rate performance for the proposed 1-bit precoding algorithm.

\end{abstract}
\begin{keywords}
Massive MIMO, 1-bit precoding, constructive interference, Lagrangian, branch-and-bound.
\end{keywords}
\section{Introduction}
Massive multiple-input multiple-output (MIMO), a.k.a. large-scale antenna arrays, has become a key enabling technique for the coming fifth-generation (5G) wireless communication systems \cite{intro1}\nocite{intro2}-\cite{intro4}. In the downlink transmission of a massive MIMO system, low-complexity linear precoding methods \cite{RZF} are shown to be near-optimal, while non-linear precoding approaches \cite{THP}\nocite{VP1}-\cite{VP2} are not preferred due to their high computational costs. Nevertheless, this near optimality is built on fully-digital signal processing with high-resolution digital-to-analog converters (DACs), while such a direct extension from small-scale antenna arrays to large-scale ones will incur prohibitive hardware complexity. The consequent power consumption at the base station (BS) will also be huge, which does not meet the target of energy-efficient transmission for future wireless communication systems. To this end, hardware-efficient large-scale antenna architectures such as hybrid analog-digital structures \cite{hybrid-1}\nocite{hybrid-2}\nocite{hybrid-3}-\cite{hybrid-4}, constant-envelope transmission \cite{cep2}\nocite{cep3}-\cite{cep4}, and low-resolution DACs have been proposed, where the use of low-resolution DACs, more specifically 1-bit DACs, is the focus of this paper.

In the literature, there already exist some works that study the precoding design in the presence of 1-bit DACs . This includes linear 1-bit precoding designs in \cite{dac1}, \cite{dac2} as well as nonlinear 1-bit precoding designs \cite{dac6}\nocite{dac7}\nocite{dac8}-\cite{dac9}, where non-linear 1-bit precoding schemes generally perform much better than linear ones. More specifically, in \cite{dac6} and \cite{dac7}, non-linear 1-bit precoding schemes were proposed via the gradient projection algorithm based on the minimum error rate metric and minimum mean-squared error (MMSE) metric, respectively. \cite{dac8} proposed a 1-bit precoding design via a biconvex relaxation procedure, while \cite{dac9} extended the work in \cite{dac8} and proposed several 1-bit precoding schemes based on semidefinite relaxation (SDR), $\ell_\infty$-norm relaxation and sphere precoding, respectively. It should be noted that the above non-linear 1-bit precoding schemes that achieve promising error rate performances operate on a symbol level, i.e., the precoding strategy and the precoded signals are designed based on both the data symbols to be transmitted and the channel state information (CSI), as opposed to many traditional precoding designs that are dependent on CSI only \cite{RZF}-\cite{VP2}.

When it comes to symbol-level precoding, there is a concept termed `constructive interference' (CI) that has already received increasing research attention in recent years \cite{ci1}, \cite{ci2}. CI is defined as the interference that pushes the received signals deeper in the decision region and farther away from the decision boundaries, which further improves the detection performance, though the MSE in this case will increase. This observation has been exploited in \cite{ci3} and the references therein by symbol-level CI precoding to achieve an improved error rate performance in traditional multi-user MIMO scenarios. Inspired by this concept, \cite{dac16}\nocite{dac12}-\cite{dac13} extended the idea of CI to 1-bit precoding designs, and the resulting performance is shown to be promising. Moreover, while not explicitly shown, \cite{dac14} also adopts the formulation of CI-based 1-bit precoding, where a branch-and-bound (BB) based 1-bit precoding algorithm that returns the optimal solution is presented. However, this 1-bit design is based on the fully-BB (F-BB) process, which is not practically useful in massive MIMO systems due to its unfavorable complexity.

In this paper, we design a near-optimal 1-bit precoding approach that aims to minimize the symbol error rate (SER) in the downlink transmission of a multi-user large-scale antenna system, where the BB framework is leveraged. By exploiting the concept of CI and adopting the `symbol-scaling' metric, the SER minimization is equivalent to the maximization on the scaling coefficients, based on which we formulate the optimization problem, which is non-convex due to the discrete 1-bit constraint on the transmit signals. By relaxing the 1-bit constraint and further analyzing the relaxed convex problem, we mathematically prove that the majority of the entries in the transmit signal vector obtained from solving the relaxed convex problem already comply with the 1-bit constraint, i.e., only a small part of the entries need to be further normalized to meet the 1-bit requirement. Building upon this observation, we introduce the proposed 1-bit precoding design based on the P-BB procedure to further improve the performance, where the BB process is only performed for the entries that do not satisfy the 1-bit constraint. Within the BB process, we employ the `max-min' criterion to design the P-BB algorithm and adopt the adaptive subdivision rule to guarantee a fast convergence speed. Compared to the traditional F-BB methods whose complexity becomes prohibitive when large-scale antenna arrays are considered, our proposed P-BB approach makes the BB framework applicable in such scenarios with significantly reduced complexity, while still exhibiting a near-optimal error rate performance, as validated by our numerical results. 

{\bf Notations:} $a$, $\bf a$, and $\bf A$ denote scalar, column vector and matrix, respectively. ${( \cdot )^\text{T}}$ and ${( \cdot )^\text{H}}$ denote transpose and conjugate transpose, respectively. $\text{card} \left( {\cdot} \right)$ denotes the cardinality of a set, ${\text {sgn}} \left[  \cdot  \right]$ is the sign function, and $\jmath$ is the imaginary unit. $\left|  \cdot  \right|$ denotes the modulus or the absolute value, and $\left\|  \cdot  \right\|_2$ denotes the $\ell_2$-norm. ${{\cal C}^{n \times n}}$ and ${{\cal R}^{n \times n}}$ represent an $n \times n$ matrix in the complex  and real set, respectively. $\Re ( \cdot )$ and $\Im ( \cdot )$ represent the extraction of the real and imaginary part, respectively, and ${\bf I}_K$ represents a $K \times K$ identity matrix.

\section{System Model}
We focus on a generic downlink massive MIMO system, where the BS with each RF chain equipped with a pair of 1-bit DACs communicates with multiple single-antenna users in the same time-frequency resource simultaneously. We denote the total number of transmit antennas at the BS by $N_t$ and the total number of users by $K$, where $N_t \gg K$. Since we focus on the effect of 1-bit DACs on the data transmission, we assume ideal ADCs are adopted at each receiver and perfect CSI is available at the BS \cite{dac1}-\cite{dac9}. We denote the intended data symbol for user $k$ by $s_k$, which is assumed to be drawn from a unit-norm $\mathbb M$-PSK constellation, and we express the data symbol vector as ${\bf s}=\left[ {s_1, s_2, \cdots, s_K} \right]^\text{T} \in {\cal C}^{K \times 1}$. We consider a flat-fading Rayleigh channel between the BS and the users, denoted by ${\bf H}=\left[ {{\bf h}_1, {\bf h}_2, \cdots, {\bf h}_K} \right]^\text{T} \in {\cal C}^{K \times N_t}$, with each entry following a standard complex Gaussian distribution $\mathbb {CN}\left( {0,1} \right)$, where ${\bf h}_k$ is the channel vector between the BS and user $k$. Accordingly, the transmit signal vector ${\bf x} \in {\cal C}^{N_t \times 1}$ at the antenna port can be expressed as
\begin{equation}
{\bf{x}} = {\cal Q}\left( {{\bf{\tilde x}}} \right) = {\cal Q}\left( {{\cal P}\left( {{\bf{s}},{\bf{H}}} \right)} \right),
\label{eq_1}
\end{equation}
where ${\bf{\tilde x}}={\cal P}\left( {{\bf{s}},{\bf{H}}} \right)$ represents the unquantized transmit signal vector, and ${\cal Q}$ is the 1-bit quantization operation. ${\cal P}$ forms the unquantized signal vector ${\bf{\tilde x}}$ based on the knowledge of $\bf s$ and $\bf H$, which represents the precoding strategy adopted at the BS. In this paper, we normalize ${\left\| {\bf{x}} \right\|_2^2} = 1$ such that each entry in $\bf x$ satisfies
\begin{equation}
{x_n} \in {{\cal X}_{{\text{DAC}}}}, {\kern 3pt}\forall n \in {\cal N},
\label{eq_2}
\end{equation}
where ${\cal N}=\left\{ {1,2,\cdots,N_t} \right\}$ and ${\cal X}_\text{DAC}=\left\{ { \pm \frac{1}{{\sqrt {2{N_t}} }} \pm \frac{1}{{\sqrt {2{N_t}} }}\jmath} \right\}$. The received signal vector ${\bf y} \in {\cal C}^{K \times 1}$ can be expressed as
\begin{equation}
{\bf y}={\bf Hx}+{\bf n},
\label{eq_3}
\end{equation}
where ${\bf n}\in {\cal C}^{K \times 1}$ is the additive Gaussian noise at the receiver side with ${\bf{n}} \sim {\mathbb {CN}}\left( {{\bf{0}},{\sigma ^2} \cdot {{\bf{I}}_K}} \right)$.

\section{Proposed 1-Bit Precoding via P-BB}

\begin{figure}[!t]
\centering
\includegraphics[scale=0.35]{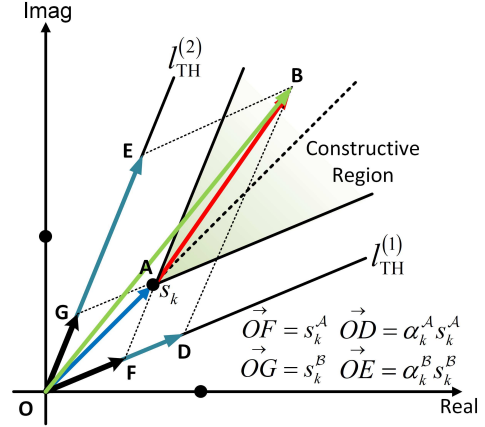}
\caption{An illustrative example of CI condition for PSK}
\end{figure}

\subsection{Problem Formulation}
We present the construction of the 1-bit precoding optimization problem based on the `symbol-scaling' CI metric in this section. The `symbol-scaling' CI formulation performs a signal decomposition of the data symbols as well as the noiseless received signals, where the introduced scaling coefficients are the variables to be optimized. To be more specific, we depict one quarter of an 8PSK constellation as the example in Fig. 1, where without loss of generality we denote the data symbol for user $k$ by $\vec{OA}=s_k$, which is further decomposed along the two detection boundaries of $s_k$ into \cite{ci3}
\begin{equation}
\vec{OA}=\vec{OF}+\vec{OG} = s_k^{\cal A} + s_k^{\cal B}, {\kern 3pt} \forall k \in {\cal K},
\label{eq_4}
\end{equation}
where ${\cal K}=\left\{ {1,2,\cdots,K} \right\}$, $\vec{OF}=s_k^{\cal A}$ and $\vec{OG}=s_k^{\cal B}$ are parallel to the detection boundary $l_\text{TH}^{\left( 1 \right)}$ and $l_\text{TH}^{\left( 2 \right)}$ respectively, as shown in Fig. 1. We refer the interested readers to Section IV of \cite{dac13} for the expression of $s_k^{\cal A}$ and $s_k^{\cal B}$ when a generic $\mathbb M$-PSK modulation is employed, which are omitted here for brevity. We further denote the received signal excluding noise for user $k$ by $\vec{OB}$, which is similarly decomposed into
\begin{equation}
\vec{OB}={\bf{h}}_k^\text{T}{\bf{x}} = \alpha _k^{\cal A} s_k^{\cal A} + \alpha _k^{\cal B} s_k^{\cal B}, {\kern 3pt} \forall k \in {\cal K},
\label{eq_5}
\end{equation}
where $\alpha _k^{\cal A} \ge 0$ and $\alpha _k^{\cal B} \ge 0$ are the introduced scaling coefficients that jointly represent the effect of interference and 1-bit quantization on $s_k$. Following \cite{ci3}, minimizing the SER is equivalent to pushing the noiseless received signal $\vec{OB}$ as deep as possible in the decision region and farther away from both of the decision boundaries, which is further equivalent to maximizing the minimum value of $\left\{ {\alpha _k^{\cal A}, \alpha _k^{\cal B}} \right\}$. Accordingly, the 1-bit precoding design can be formulated as
\begin{equation}
\begin{aligned}
&\mathcal{P}_1: {\kern 3pt} \mathop {\max }\limits_{\bf{x}} \mathop {\min }\limits_{k, {\kern 1pt} {\cal U}} {\kern 3pt} \alpha _k^{\cal U} \\
&{\kern 2pt} \text{s.t.} {\kern 10pt} {\bf C1:} {{\bf{h}}_k^\text{T}}{\bf x} = \alpha_k^{\cal A} s_k^{\cal A} + \alpha_k^{\cal B} s_k^{\cal B}, {\kern 3pt} \forall k \in {\cal K}; \\
&{\kern 22.5pt}{\bf C2:} x_n \in {\cal X}_{{\text{DAC}}}, {\kern 3pt} \forall n \in {\cal N}; {\kern 10pt} {\bf C3:} {\cal U} \in \left\{ {{\cal A}, {\cal B}} \right\},
\label{eq_6}
\end{aligned}
\end{equation}
which is a non-convex optimization problem due to the discrete 1-bit constraint $x_n \in {\cal X}_{{\text{DAC}}}$, $\forall n \in {\cal N}$.

\subsection{Analytical Study on 1-bit CI Precoding}
While ${\cal P}_1$ is originally a non-convex optimization problem and difficult to solve, by relaxing the 1-bit constraints in ${\cal P}_1$, we arrive at a convex problem formulation:
\begin{equation}
\begin{aligned}
&\mathcal{P}_2: {\kern 3pt} \mathop {\max }\limits_{\bf \tilde{x}} \mathop {\min }\limits_{k, {\kern 1pt} {\cal U}} {\kern 3pt} \alpha _k^{\cal U} \\
&{\kern 2pt} \text{s.t.} {\kern 10pt} {\bf C1:} {{\bf{h}}_k^\text{T}}{\bf \tilde x} = \alpha_k^{\cal A} s_k^{\cal A} + \alpha_k^{\cal B} s_k^{\cal B}, {\kern 3pt} \forall k \in {\cal K}; \\
&{\kern 22.5pt} {\bf C2:} \left| {\Re \left( {{{\tilde x}_n}} \right)} \right| \le \frac{1}{{\sqrt {2{N_t}} }}, {\kern 3pt} \left| {\Im \left( {{{\tilde x}_n}} \right)} \right| \le \frac{1}{{\sqrt {2{N_t}} }}, {\kern 3pt} \forall n \in {\cal N}; \\
&{\kern 22.5pt} {\bf C3:} {\cal U} \in \left\{ {{\cal A}, {\cal B}} \right\}.
\label{eq_7}
\end{aligned}
\end{equation}
A sub-optimal solution can then be obtained by enforcing the 1-bit constraint on the signal vector $\bf \tilde x$ obtained from solving ${\cal P}_2$, i.e.,
\begin{equation}
{x_n} = \frac{{{\mathop{\text {sgn}}} \left[ {\Re \left( {{{\tilde x}_n}} \right)} \right]}}{{\sqrt {2{N_t}} }} + \frac{{{\mathop{\text {sgn}}} \left[ {\Im \left( {{{\tilde x}_n}} \right)} \right]}}{{\sqrt {2{N_t}} }}\jmath, {\kern 3pt} \forall n \in {\cal N}.
\label{eq_8}
\end{equation}
We denote the above relaxation-normalization procedure by `CI 1-Bit' and the corresponding transmit signal vector by ${\bf x}_\text{CI}^\text{PSK}$.

Based on Lagrangian and KKT conditions, in this section we further elaborate on ${\cal P}_2$ and show that most of the entries in $\bf \tilde x$ from solving ${\cal P}_2$ already comply with the 1-bit requirement, as a motivation for our proposed 1-bit precoding approach via P-BB. To be more specific, we first express $\alpha_k^{\cal A}$ and $\alpha_k^{\cal B}$ as a function of $\bf x$, given by
\begin{equation}
\begin{aligned}
\alpha _k^{\cal A} =& {\kern 2pt} \frac{{\Im \left( {s_k^{\cal B}} \right)\Re \left( {{\bf{h}}_k^\text{T}} \right) - \Re \left( {s_k^{\cal B}} \right)\Im \left( {{\bf{h}}_k^\text{T}} \right)}}{{\Re \left( {s_k^{\cal A}} \right)\Im \left( {s_k^{\cal B}} \right) - \Im \left( {s_k^{\cal A}} \right)\Re \left( {s_k^{\cal B}} \right)}} \cdot \Re \left( {\bf{x}} \right)\\
& - \frac{{\Im \left( {s_k^{\cal B}} \right)\Im \left( {{\bf{h}}_k^\text{T}} \right) + \Re \left( {s_k^{\cal B}} \right)\Re \left( {{\bf{h}}_k^\text{T}} \right)}}{{\Re \left( {s_k^A} \right)\Im \left( {s_k^B} \right) - \Im \left( {s_k^{\cal A}} \right)\Re \left( {s_k^{\cal B}} \right)}} \cdot \Im \left( {\bf{x}} \right) \\
=& {\kern 2pt} {\bf{a}}_k^\text{T}\Re \left( {\bf{x}} \right) + {\bf{b}}_k^\text{T}\Im \left( {\bf{x}} \right), \\
\alpha _k^{\cal B} =& {\kern 2pt} \frac{{\Re \left( {s_k^{\cal A}} \right)\Im \left( {{\bf{h}}_k^\text{T}} \right) - \Im \left( {s_k^{\cal A}} \right)\Re \left( {{\bf{h}}_k^\text{T}} \right)}}{{\Re \left( {s_k^{\cal A}} \right)\Im \left( {s_k^{\cal B}} \right) - \Im \left( {s_k^{\cal A}} \right)\Re \left( {s_k^{\cal B}} \right)}} \cdot \Re \left( {\bf{x}} \right)\\
& + \frac{{\Re \left( {s_k^{\cal A}} \right)\Re \left( {{\bf{h}}_k^\text{T}} \right) + \Im \left( {s_k^{\cal A}} \right)\Im \left( {{\bf{h}}_k^\text{T}} \right)}}{{\Re \left( {s_k^{\cal A}} \right)\Im \left( {s_k^{\cal B}} \right) - \Im \left( {s_k^{\cal A}} \right)\Re \left( {s_k^{\cal B}} \right)}} \cdot \Im \left( {\bf{x}} \right) \\
=& {\kern 2pt} {\bf{c}}_k^\text{T}\Re \left( {\bf{x}} \right) + {\bf{d}}_k^\text{T}\Im \left( {\bf{x}} \right).
\end{aligned}
\label{eq_9}
\end{equation}
which is obtained by comparing the real and imaginary part of both sides of \eqref{eq_5}. By expressing ${{\bf{x}}_{\bf{E}}} = {\left[ {\Re \left( {{{\bf{x}}^\text{T}}} \right),\Im \left( {{{\bf{x}}^\text{T}}} \right)} \right]^\text{T}}$, ${\bf \Lambda} = {\left[ {\alpha _1^{\cal A}, \alpha _2^{\cal A}, \cdots ,\alpha _K^{\cal A},\alpha _1^{\cal B}, \alpha _2^{\cal B}, \cdots ,\alpha _K^{\cal B}} \right]^\text{T}}$, and further defining
\begin{equation}
{\bf{p}}_k^\text{T} = \left[ {{\bf{a}}_k^\text{T},{\bf{b}}_k^\text{T}} \right], {\kern 3pt} {\bf{q}}_k^\text{T} = \left[ {{\bf{c}}_k^\text{T}, {\kern 3pt} {\bf{d}}_k^\text{T}} \right], \forall k \in {\cal K},
\label{eq_10}
\end{equation}
\eqref{eq_9} can be expressed in a compact matrix form as
\begin{equation}
{\bf \Lambda}={\bf M}{\bf x_E},
\label{eq_11}
\end{equation}
where ${\bf M} \in {\cal R}^{2K \times 2N_t}$ is given by
\begin{equation}
{\bf{M}} = {\left[ {{\bf{p}}_1,{\bf{p}}_2, \cdots ,{\bf{p}}_K,{\bf{q}}_1,{\bf{q}}_2, \cdots ,{\bf{q}}_K} \right]^\text{T}}.
\label{eq_12}
\end{equation}
Based on this transformation, ${\cal P}_2$ is equivalent to the following optimization problem:
\begin{equation}
\begin{aligned}
&\mathcal{P}_3: {\kern 3pt} \mathop {\max }\limits_{{\bf \tilde x}_{\bf E}} \mathop {\min }\limits_{l} {\kern 3pt} \alpha _l \\
&{\kern 2pt} \text{s.t.} {\kern 10pt} {\bf C1:} \alpha _l= {\bf m}_l^\text{T}{\bf \tilde x}_{\bf E}, {\kern 3pt} \forall l \in {\cal L}; \\
&{\kern 22.5pt} {\bf C2:} \left| {{{\tilde x}_m^{\text E}}} \right| \le \frac{1}{{\sqrt {2{N_t}} }}, {\kern 3pt} \forall m \in {\cal M},
\label{eq_13}
\end{aligned}
\end{equation}
where ${\bf m}_l^\text{T}$ represents the $l$-th row in $\bf M$, ${{\tilde x}_n^{\text E}}$ is the $n$-th entry in ${\bf \tilde x}_{\bf E}$, ${\cal L}=\left\{ {1,2,\cdots,2K} \right\}$, and ${\cal M}=\left\{ {1,2,\cdots,2N_t} \right\}$. Based on the formulation of ${\cal P}_3$, we derive the following important proposition, which builds the foundation of the proposed P-BB procedure in the following.

{\bf Proposition:} For $\bf \tilde x_E$ obtained by solving ${\cal P}_3$, there are at least a total number of $\left( {2{N_t} - 2K + 1} \right)$ entries that already comply with the 1-bit requirement.

{\bf Proof:} Proving this proposition is equivalent to proving that there are at most a total number of $\left( {2K - 1} \right)$ entries in $\bf \tilde x_E$ whose amplitudes are smaller than $\frac{1}{\sqrt{2N_t}}$.

To begin with, by transforming ${\cal P}_3$ into a standard minimization form: 
\begin{equation}
\begin{aligned}
&\mathcal{P}_{4}: {\kern 3pt} \mathop {\min }\limits_{t,{{{\bf{\tilde x}}}_{\bf{E}}}} {\kern 3pt} -t \\
&{\kern 2pt} \text{s.t.} {\kern 10pt} {\bf C1:} t - {\bf{m}}_l^\text{T}{{\bf \tilde x}_{\bf{E}}} \le 0, {\kern 3pt} \forall l \in {\cal L}; \\
&{\kern 22.5pt} {\bf C2:} \tilde x_m^{\text E} - \frac{1}{{\sqrt {2{N_t}} }} \le 0, {\kern 3pt} -\tilde x_m^{\text E} - \frac{1}{{\sqrt {2{N_t}} }} \le 0, {\kern 3pt} \forall m \in {\cal M},
\label{eq_14}
\end{aligned}
\end{equation}
we express the Lagrangian of ${\cal P}_4$ as
\begin{equation}
\begin{aligned}
&{\cal L}\left( {t,{{{\bf{\tilde x}}}_{\bf{E}}},{\beta _l},{\mu _m},{\nu _m}} \right) =  - t + \sum\limits_{l = 1}^{2K} {{\beta _l}\left( {t - {\bf{m}}_l^\text{T}{{\bf \tilde {x}}_{\bf{E}}}} \right)} \\
& {\kern 20pt} + \sum\limits_{m = 1}^{2{N_t}} {{\mu _m}\left( {\tilde x_m^{\text E} - \frac{1}{{\sqrt {2{N_t}} }}} \right)}  - \sum\limits_{m = 1}^{2{N_t}} {{\nu _m}\left( {\tilde x_m^{\text E} + \frac{1}{{\sqrt {2{N_t}} }}} \right)}\\
&=\left( {{{\bf{1}}^\text{T}}{\bm \beta}  - 1} \right)t - {{\bm \beta} ^\text{T}}{\bf{M}}{{\bf{\tilde x}}_{\bf{E}}} + \left( {{{\bm \mu} ^\text{T}} - {{\bm \nu} ^\text{T}}} \right){{\bf{\tilde x}}_{\bf{E}}} \\
& {\kern 12pt} - \frac{1}{{\sqrt {2{N_t}} }}\left( {{{\bf{1}}^\text{T}}{\bm \mu}  + {{\bf{1}}^\text{T}}{\bm \nu} } \right),
\end{aligned}
\label{eq_15}
\end{equation}
where ${\beta _l}$, ${\mu _m}$, and ${\nu _m}$ are the non-negative Lagrangian multipliers. We then construct the KKT conditions as:
\begin{IEEEeqnarray}{rCl} 
\IEEEyesnumber
\frac{{\partial {\cal L}}}{{\partial t}} =  {{\bf{1}}^\text{T}}{\bm \beta}  - 1  = 0 {\kern 30pt} \IEEEyessubnumber* \label{eq_16a} \\
\frac{{\partial {\cal L}}}{{\partial {{{\bf{\tilde x}}}_{\bf{E}}}}} =  - {{\bf{M}}^\text{T}}{\bm \beta}  + {\bm \mu}  - {\bm \nu}  = {\bf{0}} {\kern 30pt} \label{eq_16b} \\
{{\beta _l}\left( {t - {\bf{m}}_l^\text{T}{{\bf \tilde {x}}_{\bf{E}}}} \right)}=0, {\kern 3pt} \beta_l \ge 0, {\kern 3pt} \forall l \in {\cal L} {\kern 30pt} \label{eq_16c} \\
{{\mu _m}\left( {\tilde x_m^{\text E} - \frac{1}{{\sqrt {2{N_t}} }}} \right)}=0, {\kern 3pt} {\mu _m}\ge 0, {\kern 3pt} \forall m \in {\cal M} {\kern 30pt} \label{eq_16d} \\
{{\nu _m}\left( {\tilde x_m^{\text E} + \frac{1}{{\sqrt {2{N_t}} }}} \right)}=0, {\kern 3pt} {\nu _m}\ge 0, {\kern 3pt} \forall m \in {\cal M} {\kern 30pt} \label{eq_16e}
\end{IEEEeqnarray}
In the following, we prove this proposition by contradiction.

Suppose that there are a total number of $2K$ entries in ${\bf \tilde x}_{\bf E}$ whose amplitudes are strictly smaller than $\frac{1}{\sqrt {2N_t}}$, and for notational convenience we introduce a set $\cal S$ to include the indices of these entries, which is mathematically expressed as
\begin{equation}
n \in {\cal S}, {\kern 3pt} {\text {if}} {\kern 3pt} \left| {\tilde x_n^{\text E}} \right| < \frac{1}{{\sqrt {2{N_t}} }},
\label{eq_17}
\end{equation}
where we have $\text{card}\left( {\cal S} \right)=2K$ based on our above assumption. According to the complementary slackness conditions \eqref{eq_16d} and \eqref{eq_16e}, we obtain
\begin{equation}
\mu _n=0, {\kern 3pt} \nu_n=0, {\kern 3pt} \forall n \in {\cal S}.
\label{eq_18}
\end{equation}
Recall \eqref{eq_16b} which can be regarded as a linear equation with $\bm \beta$ as the variable, and for simplicity we introduce ${\bf W}={\bf M}^\text{T}={\left[ {{\bf{w}}_1,{\bf{w}}_2, \cdots ,{\bf{w}}_{2{N_t}}} \right]^\text{T}}$. Given \eqref{eq_18}, we subsequently pick the corresponding rows of ${\bf W}$ whose indices belong to $\cal S$ to formulate a sub linear equation:
\begin{equation}
{\bf W_p}{\bm \beta}  = \hat {\bm \mu}_{\bf p}  - \hat {\bm \nu}_{\bf p}  = {\bf{0}},
\label{eq_19}
\end{equation}
where ${\bf W_p} \in {\cal R}^{\text{card}\left( {\cal S} \right) \times 2K}$ is expressed as
\begin{equation}
{\bf W_p} = {\left[ {{\bf{w}}_{{n_1}}, \cdots ,{\bf{w}}_{{n_m}}, \cdots ,{\bf{w}}_{{n_{\text{card}\left( {\cal S} \right)}}}} \right]^\text{T}}, {\kern 3pt} \forall n_m \in {\cal S}.
\label{eq_20}
\end{equation}
Based on that $\text{card}\left( {\cal S}\right) = 2K$, we obtain that ${\bf W_p}$ is full-rank. According to the linear algebra theory \cite{book1}, given a full-rank coefficient matrix $\bf W_p$, a non-zero solution to \eqref{eq_19} does not exist and there is only a trivial solution, i.e.,
\begin{equation}
{\bm \beta}^*={\bf 0}.
\label{eq_21}
\end{equation}
However, this solution does not comply with \eqref{eq_16a} that enforces a non-zero solution of $\bm \beta$, which causes contradiction. By following a step similar to the above, this contradiction is also observed if we assume there are a total number of $N>2K$ entries in the obtained $\bf \tilde x_E$ whose amplitudes are strictly smaller than $\frac{1}{\sqrt{2N_t}}$, which completes the proof.${\kern 203pt} \blacksquare$

\subsection{Proposed 1-Bit Precoding via Partial Branch-and-Bound}
Based on the results in the {\bf Proposition}, we propose the 1-bit precoding design via P-BB in this section, which essentially performs the BB process for part of the entries only, more specifically the entries in $\bf \tilde x_E$ that do not comply with the 1-bit requirement, as opposed to traditional BB-based schemes that perform BB process for all the entries in the transmit signal vector. This allows a considerable complexity reduction while still exhibiting a near-optimal SER performance, as will be shown by the numerical results.

To begin with, we perform row rearrangements on $\bf \tilde x_E$ to arrive at $\bf \hat x_E$, such that it can be decomposed into
\begin{equation}
{{\bf{\hat x}}_{\bf{E}}} = {\left[ {{\bf{x}}_{\bf{F}}^\text{T},{\bf{x}}_{\bf{R}}^\text{T}} \right]^\text{T}},
\label{eq_22}
\end{equation}
where ${\bf{x}}_{\bf{F}} \in {\cal R}^{N_F \times 1}$ consists of $x_m^{\text E}$ that already satisfy the 1-bit constraint and is fixed throughout the P-BB procedure. ${\bf {x}}_{\bf{R}}=\left[ {x_1^{\text R},x_2^{\text R}, \cdots ,x_{N_R}^{\text R}} \right]^\text{T}$ consists of the residual entries in ${{\bf{\hat x}}_{\bf{E}}}$ whose amplitudes are smaller than $\frac{1}{\sqrt {2N_t}}$. Following the {\bf Proposition}, we obtain $N_F \ge \left({2N_t-2K+1}\right)$, $N_R \le \left({2K-1}\right)$ and $N_F+N_R=2N_t$. Similarly, we rearrange $\bf M$ into $\bf \hat M$ such that ${\bf \hat M}{\bf \hat x_E}={\bf M}{\bf \tilde x_E}$, which is also decomposed into
\begin{equation}
{\bf \hat{M}} = \left[ {{{\bf{M}}_{\bf{F}}},{{\bf{M}}_{\bf{R}}}} \right],
\label{eq_23}
\end{equation}
where ${{\bf{M}}_{\bf{F}}} = {\left[ {{\bf \hat{m}}_1^\text{F},{\bf \hat{m}}_2^\text{F}, \cdots ,{\bf \hat{m}}_{2K}^\text{F}} \right]^\text{T}} \in {\cal R}^{2K\times N_F}$ and ${{\bf{M}}_{\bf{R}}} = {\left[ {{\bf \hat{m}}_1^\text{R},{\bf \hat{m}}_2^\text{R}, \cdots ,{\bf \hat{m}}_{2K}^\text{R}} \right]^\text{T}} \in {\cal R}^{2K\times N_R}$. The proposed P-BB approach aims to further optimize $\bf x_R$ with $\bf x_F$ fixed, which leads to the following optimization problem:
\begin{equation}
\begin{aligned}
&\mathcal{P}_5: {\kern 3pt} \mathop {\min }\limits_{{{\bf{x}}_{\bf{R}}}} {\kern 3pt} - t \\
&{\kern 2pt} \text{s.t.} {\kern 10pt} {\bf C1:} t - {\left( {{\bf{\hat m}}_l^\text{R}} \right)^\text{T}}{{\bf{x}}_{\bf{R}}} \le {\left( {{\bf{\hat m}}_l^\text{F}} \right)^\text{T}}{{\bf{x}}_{\bf{F}}}, {\kern 3pt} \forall l \in {\cal L}; \\
&{\kern 22.5pt} {\bf C2:} {x_m^{\text R}} \in {\cal X}_\text{DAC}, {\kern 3pt} \forall m = \left\{ {1,2,\cdots, N_R} \right\}.
\label{eq_24}
\end{aligned}
\end{equation}
The subsequent BB procedure follows \cite{dac14} and is omitted here due to the limited space, where we note that to guarantee a fast convergence speed, in the branching process we adopt the adaptive subdivision rule to choose the index of the entry in ${\bf x}_{\bf R}$ which is to be allocated a value in the current iteration. The corresponding index of the entry that is chosen should satisfy:
\begin{equation}
n= \arg \mathop {\max }\limits_n \left| {x_n^\text{R} - {\cal Q}\left( {x_n^\text{R}} \right)} \right|,
\label{eq_25}
\end{equation}
where $x_n^\text{R}$ is the $n$-th entry in $\bf x_R$ and $n$ is its corresponding index. This proposed 1-bit precoding algorithm is termed `CI 1-Bit P-BB'.

\section{Numerical Results}
We present numerical results of the proposed 1-bit precoding design in this section based on Monte Carlo simulations. In each figure, we define the transmit SNR as $\rho=\frac{1}{\sigma^2}$ by assuming unit transmit power, and we compare our proposed P-BB based scheme with both linear and non-linear 1-bit precoding designs in the literature.

\begin{figure}[!t]
\centering
\includegraphics[scale=0.31]{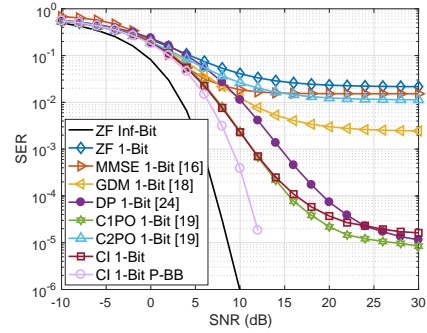}
\caption{SER v.s. transmit SNR, QPSK, $N_t=64$, $K=16$}
\end{figure}

In Fig. 2, we present the SER result for QPSK modulation in a $64\times16$ MIMO system. Compared to existing 1-bit precoding algorithms, our proposed 1-bit precoding via P-BB achieves a noticeable improvement in terms of the error rate and eliminates the error floor that are commonly observed for traditional 1-bit precoding methods when the SNR becomes high. A similar trend is observed when we extend the modulation type to 8PSK, as depicted in Fig. 3 for a $128\times16$ MIMO system. Both the above results validate the effectiveness of the proposed 1-bit precoding approach via P-BB.

\begin{figure}[!t]
\centering
\includegraphics[scale=0.31]{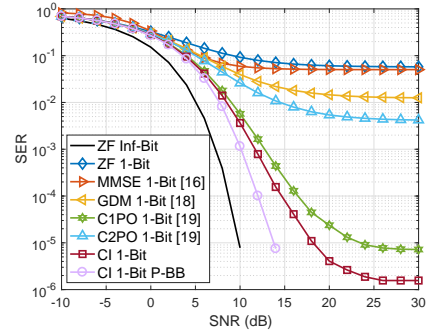}
\caption{SER v.s. transmit SNR, 8PSK, $N_t=128$, $K=16$}
\end{figure}

\section{Conclusion}
In this paper, we have proposed a 1-bit precoding approach via the P-BB procedure, which significantly outperforms existing 1-bit precoding schemes and is shown to achieve near-optimal error rate performance. The proposed 1-bit precoding scheme is built on the observation that most of the entries in the obtained transmit signal vector already satisfy the 1-bit requirement by solving the relaxed 1-bit precoding problem, and thus the BB process is only needed for the residual entries that do not comply with the 1-bit constraint. The proposed 1-bit precoding scheme also enables the BB framework to be applicable in large-scale antenna arrays, which was not applicable due to the prohibitive complexity. 


\bibliographystyle{IEEEbib}
\bibliography{refs}

\end{document}